\documentclass[11pt]{article}

\pdfoutput=1
\usepackage{tikz}
\usepackage{subfigure}
\usepackage{amsmath}
\usepackage{amssymb}
\usepackage{graphics}
\usepackage{rotating}
\usepackage{array}
\usepackage{graphicx}
\usepackage{epstopdf}
\usepackage{hyperref}
\usepackage{xcolor}
\renewcommand\[{\begin{equation}}  
\renewcommand\]{\end{equation}}

\renewcommand{\d}{{\rm{d}}}
\renewcommand{\H}{{\cal H}}
\newcommand{\ee}{{\rm{e}}}
\newcommand{\eps}{{\epsilon}}
\newcommand{\dt}{{\rm{d} t}}
\newcommand{\bol}[1]{{\bf #1}}

\numberwithin{equation}{section}

\begin{document}

\title{Simulating rare events in dynamical processes}

\author{
Cristian Giardina\footnote{Universit\`a di Modena e Reggio Emilia, viale A. Allegri, 9 - 42121 Reggio Emilia , Italy} \and
Jorge Kurchan\footnote{ESPCI, 10 rue Vauquelin, Paris, France 75005 - CNRS UMR 7636 PMMH} \and
Vivien Lecomte  \footnote{Univ Paris Diderot, Sorbonne Paris Cite, LPMA, UMR 7599 CNRS, F75205 Paris, France} \and  
Julien Tailleur \footnote{Univ Paris Diderot, Sorbonne Paris Cite, MSC, UMR 7057 CNRS, F75205 Paris, France}
}

\maketitle 

\renewcommand{\baselinestretch}{1.2}\large

\begin{abstract}
Atypical, rare trajectories of dynamical systems are 
important: they are often the paths for chemical reactions, the haven
of (relative) stability of planetary systems, the rogue waves that are
detected in oil platforms, the structures that are responsible for
intermittency in a turbulent liquid, the active regions that allow a
supercooled liquid to flow...  Simulating them in an efficient,
accelerated way, is in fact quite simple.

In this paper we review a computational 
technique to study such rare events in both stochastic and Hamiltonian systems.
The method is based on the evolution of a family of copies of the system
which are replicated or killed in such a way as to favor the realization of the 
atypical trajectories. We illustrate this with various examples.

\end{abstract}

\section{Introduction}

When a {dynamical} system is complex enough, it becomes no longer
feasible -- or indeed, interesting -- to describe every possible
trajectory. A first step is then to study what a `typical trajectory'
does.  For Hamiltonian dynamics, Statistical Mechanics provides us
with powerful techniques to compute some properties of such typical
trajectories, but for generic dynamics we must in most cases resort to
simulations.

There are many situations in which the trajectories that matter are not
the typical ones, but rather `rare' ones reached from exceptional
initial conditions, or particularly infrequently. Consider the
following examples:

$\bullet$ Planetary systems are in general chaotic, and the different sets of present 
conditions,   falling    within the range of observational error,  may lead  to widely
varying inferences  about the past and future.
Because we do not expect that an observed system has been created
recently, or will be destroyed immediately, we must understand how
this comes about, and we are naturally led to a statistical study of
the trajectories.
 
$\bullet$ Molecular dynamics is in many cases characterized by long
periods of vibrations around a local {metastable} configuration,
punctuated by relatively rapid but infrequent `activation' events,
leading to a major rearrangement. Because they are the essential steps
of chemical transformations, it is of the greatest importance to be
able to simulate such events in an accelerated way, without having to
wait for them to happen spontaneously.  There is a vast literature on
this subject.
 
$\bullet$ In a similar fashion, supercooled liquids and glasses are
characterized by vibrational dynamics, with events localized in time
{\em and space} where the transformations take place.  These `dynamic
heterogeneities' are {the analogues of reaction} paths in
chemical systems.

$\bullet$ It has long be known that, in a liquid undergoing fully
developed turbulence, due to the presence of abnormally large
fluctuations of velocities, the dynamics are  
intermittent.  The natural question is which dynamic features are
responsible for this.

$\bullet$ In the sea there have been reports of (`rogue') waves of
exceptionally large amplitudes.  They are rare, but much more common
than one would expect from a Gaussian distribution.  The subject is of
obvious interest, and is still very much open.

$\bullet$ Transport of energy or particles across a sample is
facilitated by exceptional `ballistic' trajectories, or hindered by
situations resembling traffic jams. 
  
$\bullet$ When a system is subject to external forcing, the power
injected (or the entropy production) during a given time is a quantity
that depends on the particular trajectory it is following. The Second
Law of thermodynamics sets limits on the expectation value of these
quantities, but does not limit the extent of the (rare) fluctuations.
Thus, one may extract work from a system while lowering the total
entropy, but the probability of this goes down exponentially with its
size, and with the interval of time.

All of these problems may be studied by simulating repeatedly, or for
long times, the true dynamics. However, as one may imagine, this
procedure soon becomes unfeasible.  There are basically two types of
methods to generate in a controlled way rare events.  The
{\em Path-sampling} method amounts to Monte Carlo dynamics  in trajectory space, correctly
 designed to weigh each trajectory
with the desired bias.  {A second strategy works directly in configuration space:
one introduces a
  population of copies of the initial system and relies on a mixture
  including   the original dynamics,  supplemented with a  `Darwinian pressure' --
  again, in a controlled way-- to favor the exploration of atypical
  trajectories. In this review we concentrate on the second class.}
  

The paper is organized as follows. The population dynamics with cloning
is introduced in Section \ref{pd}, where it is shown how it can be used to
compute the large deviation function (or rather its Legendre transform)
of extensive observables of the trajectories of a diffusive
dynamics with drift and a multiplicative (cloning) term. 
The relative weight of the drift and cloning terms is analyzed in
section \ref{bias}, where it is shown how a change of bases can
help in adjusting their relative contribution.
Then a series of examples from different contexts follows.
Purely stochastic systems are studied in sections
\ref{transport} and \ref{sec:activity}, where the large
deviations of, respectively, the current in 
interacting particle systems and
the dynamical activity in kinetically constrained
models are analyzed. 
Sections \ref{chaos} and \ref{gallavotti-cohen} consider
examples of deterministic dynamics, such as 
the standard map and the Hamiltonian Fermi-Pasta-Ulam
model, for which trajectory with large or small Lyapunov exponent
are studied, or the Sinai billiard, for which the symmetry associated
with the fluctuation theorem is easily verified.
The last Section \ref{planets} suggests how the numerical
method of cloning could be used also in the study
of the stability of planetary systems.


\section{Population dynamics}
\label{pd}
  
To fix ideas, consider a noisy dynamics for a vector ${\bf x}$ whose
components evolve as:
  \begin{equation} 
  { \dot x}_i = f_i({\bf x}) +\eta_i(t)
  \label{un}
  \end{equation}
with $\eta_i$ a noise which for simplicity we shall suppose is
Gaussian and white, with variance $2T_i$.  The probability of a
trajectory up to time $t$ is found by writing $\eta_i= { \dot x}_i - f_i({\bf x})$ :
  \begin{equation}
  P[  {\bf  x}(t)] \propto  e^{-\sum_i \int_o^t dt' \;\frac{ ({ \dot x}_i - f_i )^2}{4T_i}}
   \label{do}
  \end{equation}
As an example, we wish to calculate the probability that a certain
quantity $A[ {\bf x}]$ takes  a {\em time-averaged} value ${\cal{A}}_o$:
   \begin{equation}
  p({\cal{A}}_o)= \int D[  {\bf  x}]  P[  {\bf  x}(t)]   \delta \left\{ \int_o^t  dt' \;  A({\bf x}) -t {\cal{A}}_o\right\}        
   \label{tr}
  \end{equation} 
   It is more practical to compute the Laplace transform:
   \begin{eqnarray}
Z_t(\alpha)& = & \int d{\cal{A}}_o\;   p({\cal{A}}_o) e^{\alpha t {\cal{A}}_o}  =  \int D[  {\bf  x}]  P[  {\bf  x}(t)]  e^{ \left\{ \alpha \int_o^t  dt' \;  A({\bf x}) \right\}}   \nonumber \\
& & \qquad     \propto  \int D[  {\bf  x}] e^{-\sum_i \int_o^t dt' \; \frac{ ({ \dot x}_i - f_i )^2}{4T_i} +\alpha \int_o^t  dt' \;  A({\bf x}) }   
   \label{cu}
  \end{eqnarray} 
  
   In particular, for large times $ p({\cal{A}}_o)$ becomes a peaked
   function $p({\cal{A}}_o) \sim e^{-t I({\cal{A}}_o)}$, with $I({\cal
     A}_o)$ the large deviation function given by the Legendre
   transform~\cite{Touchette}:
\begin{equation}
I({\cal A}_o)  =  \sup_{\alpha} \left[ {\cal{A}}_o\alpha - \lim_{t\to\infty}\frac{1}{t}\log Z_t(\alpha)\right]
\end{equation}
  The last of equations (\ref{cu}) may be interpreted as a sum over paths with a modified weight, and may be simulated
  with path sampling methods. The strategy we describe in this paper is instead to notice that Eq. (\ref{cu}) may be interpreted as 
  describing the following  dynamics:
  
  \begin{itemize}
  
  \item Consider a population of infinitely many non-interacting
    'clones' of the system ${\bf x}^a (t)$ satisfying the original dynamics ${\bf
      \dot{x}}^{a} (t) = {\bf f}({\bf x}^a) +$ {\boldmath
      $\eta$}$^a$.  The noise of each clone is independent
      from the others.
 
    \item {At each time interval $\delta t$, each clone is either killed
    or replicated,  so that it is replaced on average by $\exp(\alpha
    A({\bf x}^a)\; \delta t)$ clones.}
  
  \end{itemize}
  
This population dynamics is such that the average cloning or pruning
rate of clones yields at large times $Z_t(\alpha)$.  In practice, we
do not simulate infinitely many clones of the initial system and we
explain in the following how to adapt the dynamics to work with a
large, but finite, {\em fixed} number of clones (typically in the
hundreds).  We shall see how this simple idea, originally applied in
the context of Diffusion Monte Carlo \cite{DMC}, may be adapted to a
number of different problems. The actual specific form of the
population dynamics involved depends on the nature of the problem
(continuous or discrete state space, continuous or discrete time,
etc): we shall specify this in each example below. Similar strategies
to simulate rare events have been advocated in other context with
great success, see for example \cite{Aldous,Grassberger,DelMoral}.

We have mentioned  so far  large deviations of a
quantity of the form:
\begin{equation}
        F[{\bf x}(t)] = \int_o^t dt' \; A({\bf x}(t'))
\label{F}
\end{equation}
In many cases, the functionals $F$ depend also on the time-derivatives
$\frac{d{\bf x}}{dt}$, and even are functions that
are non-local in time. In these cases, the cloning rate at time $t$
depends as well on the configurations at time $t'<t$.
    
The algorithms presented in this review give not only access to large deviations of the observable $F$ but 
also allow one to compute  the
average of any observable among the corresponding, atypical, histories weighted by $e^{\alpha F}$,
allowing to answer questions such as {\em ``what happens with the vorticity  of a fluid at a time and place where energy dissipation is unusually large?'' }

The average of an observable ${\mathcal O}$ at  the final time $t$
\begin{equation}
  \overline {\mathcal O}(\alpha,t) = \frac{\langle e^{\alpha F} \mathcal O(x(t)) \rangle}{\langle e^{\alpha F}\rangle}
  \label{intermediate}
\end{equation}
is recovered from the corresponding average among the
clones at  that time. The averages at intermediate times  (for $0\ll t'\ll t$) $ \overline {\mathcal O}(\alpha,t') =
\frac{\langle e^{\alpha F} \mathcal O(x(t')) \rangle}{\langle
  e^{\alpha F}\rangle} $ may also be recovered by attaching to each
clone at time $t'$ the observed value of $\mathcal O$,  and then
constructing the average $ \overline {\mathcal O}(\alpha,t')$ among
the clones which have survived until the final time $t$.  In the large
time limit $t\to\infty$, this average is not sensitive to the
precise value of $t'$ and a better sampling is achieved by attaching
to each clone the average value of ${\mathcal O}$ around time
$t'$~\cite{GKP,GJLPvDvW2009,LTreview}.

\section{Biasing the stationary distribution: drift versus cloning}
\label{bias}

Equation (\ref{cu}) is nothing but the path-integral representation of the
equation:

\begin{equation}
    \frac{dP}{dt} = - H_\alpha P
    \label{eq1}
\end{equation}
with $P({\bf x}) $ the probability distribution, and:
\begin{equation}
  H_\alpha = {-}\sum_i  T_i  \frac{\partial^2}{\partial x_i^2}
{+} \sum_i  \frac{\partial}{\partial x_i} f_i - \alpha A
    \label{eq2}
\end{equation}
The three terms in $H_\alpha$ correspond to {\em diffusion, drift}, and
{\em cloning}, respectively.

The technique of {\em dynamic importance sampling} can always be used
to reshuffle the importance of drift and cloning.  It is implemented
by making a change of basis:
\begin{equation}
    \tilde{H}_\alpha = e^{\phi({\bf x})} H_\alpha e^{-\phi({\bf x})} =
\sum_i - T_i  \frac{\partial^2}{\partial x_i^2}  +\sum_i 
\frac{\partial}{\partial x_i} \tilde{f}_i - \tilde{A}
   \label{modi}
\end{equation}
with:
\begin{eqnarray}
    \tilde{f}_i &=& f_i + 2 T_i \frac{\partial \phi}{\partial x_i}
    \nonumber \\ \tilde{A} &=& 
     \sum_i \left[ T_i\left(\frac{\partial \phi}{\partial
      x_i}\right)^2 + \frac{\partial \phi}{\partial x_i} f_i + T_i
    \frac{\partial^2 \phi}{\partial x_i^2}\right] + \alpha A  {}
    \label{modi1}
\end{eqnarray}
In general, there is not an optimal choice for the field $\phi$.
We will see examples later in different contexts.
Another way to 
understand  (\ref{modi}) is to consider the dynamics (\ref{cu}) with a modified  large deviation function:
\begin{equation}
A \rightarrow A + \frac{d \phi}{dt}  \;\;\; ; \;\;\; F = \int_0^t A(t') \; dt' + \phi(t) - \phi(0)
\label{modi2}
\end{equation}
Writing $\frac{d \phi}{dt}  = \sum_i \frac{\partial  \phi}{\partial x_i } \dot x_i$ and expressing $ \dot x_i$ in terms of the equation of motion,
we recover the result (\ref{modi}), (\ref{modi1}). Alternatively,  we may of course always consider the modified dynamics as 
the original one with a cloning rate $  A + \frac{d \phi}{dt} $.

{Trajectories are thus
reweighted according to initial and final configurations.} 
The many-time  expectation with respect to the original dynamics
$\langle O(t_1) O(t_2) ... O(t_n) \rangle $ for
$t_1<t_2<\ldots<t_n$ , starting from a distribution $P_o({\bf x})$,
corresponds to averages with the modified dynamics of $\langle O(t_1)
O(t_2) ...  O(t_n)e^{\phi(t_n)} \rangle $, starting from a
distribution $e^{\phi}P_o({\bf x})$.

 It is important to realize that
this is {\em not} the usual Monte-Carlo  importance sampling technique used in
equilibrium simulations, which consists simply of modifying the energy
 in the sampling protocol
$E \rightarrow E+B$ (for some suitably chosen $B$), and compensating
by calculating averages as follows:
\begin{equation}
     \langle O \rangle_E \rightarrow \langle O e^{\beta B}
     \rangle_{E+B}
     \label{is}
\end{equation}
where $\langle \bullet \rangle_E $ stands for average using a Monte Carlo scheme with energy $E$. 
With such a technique, one cannot calculate many-time correlation
functions, or trajectory probabilities, since the dynamics are
unrelated to the original ones; as one can see easily for the case
$B=-E$ where the modified dynamics are simple diffusion, unlike the
original ones.
In out of equilibrium  situations, we do not have an explicit
expression for the stationary distribution,
 and there is no simple way to modify the dynamics in order that they remain
probability conserving and have a biased measure,
 i.e. there is no analog of (\ref{is}). 

\subsection{Computing large moments of instantaneous \\ quantities: the example of 
turbulence.}

It sometimes happens that we are interested in calculating the moments of
an {\em instantaneous} quantity. Consider for example
the case of Navier-Stokes equations for driven turbulence. A set of quantities  that characterize 
intermittency are  the so-called  longitudinal-structure functions \cite{Frisch}
\begin{equation}
S_p(R) =  \langle  | {\bf v}({\bf x}+ {\bf R}) -    {\bf v}({\bf x}) |^p \rangle = \langle e^{p \ln    | {\bf v}({\bf x}+ {\bf R}) -    {\bf v}({\bf x}) |} \rangle
\end{equation}
In order to compute these moments efficiently, we  put, in the notation of the previous paragraphs:
\begin{equation}
\phi = \frac{p}{2}  \ln  | {\bf v}({\bf x}+ {\bf R}) -    {\bf v}({\bf x}) |^2   
\label{FF}
\end{equation}
We may run several parallel 
 simulations of fully developed turbulence in the stationary state, each with its own realization
of stochastic stirring, and supplement this with a  cloning/pruning rate equal to the time-derivative of (\ref{FF}), which may be   
expressed in terms of the instantaneous velocities using  the Navier-Stokes equations.
The total average cloning rate yields, for large times,  $S_p(R)$. 
Perhaps more interestingly,  the configurations that dominate the modified dynamics 
are the ones that   contribute to $S_p(R)$, and are continuously being sampled.
To the best of our knowledge, this strategy has not been implemented yet.

\section{Transport}
\label{transport}

We now  describe  large deviations in 
non-equilibrium stochastic models of transport.
In such models the main observables (e.g. the current,
the density, etc.)  are functions of the sample path 
of a Markov chain in a high-dimensional state space. 

\subsection{Discrete-time Markov chains}
\label{subsec:discretetimecloning}

Imagine a discretization in space of the noisy dynamics (\ref{un}), so that the 
phase space is given by a finite set of configurations. 
If we assume that also time is discretized then the dynamics can be 
described by a Markov chain $\{{\bf x}_n\}$ with  $(n=1,2,..., t)$ .
The evolution is specified by a transition probability matrix whose elements are
$p(x,y) = P({\bf x}_{n+1}=y | {\bf x}_n = x)$ and by an initial distribution $P(y) = P({\bf x}_0 = y)$. 
We consider a functional
$F[{\bf x}_n]$ which is the sum of the local contributions to the current,
an additive function of the transitions along the trajectory
up to time $t$:
\begin{equation}
F = F({\bf x}_0,{\bf x}_1,\ldots,{\bf x}_t) = \sum_{n=1}^{t} f({\bf x}_{n-1},{\bf x}_n)
\label{questa}
\end{equation}
Note that $f$ is, unlike the example in the introduction, a function
of the position at {\em two} successive times. For instance if one
considers particles diffusing on a one dimensional lattice and chooses
$f({\bf x}_{n-1},{\bf x}_n)$ to be $\pm 1$ depending on whether
particles jump to the right or the left, $F$ is the
time-integrated current flowing through the system from left to right.
The 'partition function' (\ref{cu}) is given by
\begin{eqnarray}
Z_t(\alpha) 
& = & 
\langle e^{\alpha F({\bf x}_0,{\bf x}_1,\ldots,{\bf x}_t)}\rangle \\
& = &  
\sum_{x_0,x_1,\ldots x_t} P(x_0)p(x_0,x_1)\cdots p(x_{t-1},x_{t})
e^{\alpha f(x_0,x_1)}  \cdots e^{\alpha f(x_{t-1},x_t)} 
\nonumber
\end{eqnarray} 
Just as in the previous section, we replace the initial evolution, given
by a transition matrix $p(x,y)$,  by a new evolution, given
by a matrix $p(x,y)e^{\alpha f(x,y)} $.   
We may decompose this as a probability conserving transition 
matrix \cite{GKP}:
\begin{equation}
p_\alpha(x,y) = p(x,y) e^{\alpha f(x,y)} \frac{1}{k(x)}
\label{just}
\end{equation} 
and a cloning factor
\begin{equation}
k(x) = \sum_{y} p(x,y) e^{\alpha f(x,y)}\;.
\label{justk}
\end{equation}
We  then have 
\begin{equation}
\label{multi}
Z_t(\alpha) 
=
\sum_{x_0,x_1,\ldots x_{t-1}} P(x_0)p_\alpha(x_0,x_1)\cdots p_\alpha(x_{t-2},x_{t-1})
k(x_0)\cdots k(x_{t-1}) 
\end{equation}
The convenient way to simulate (\ref{multi}) is to consider a cloning
step of average factor $k(x)$ followed by an evolution step with the
transition matrix $p_\alpha(x,y)$. The former may by implemented by
substituting a given configuration by a number $(0,1,2,...)$ of equal
clones, with expectation value of the number equal to $k(x)$, while
the latter is a transition with probability
$p_\alpha(x,y)$\footnote{
The evolution step can be easily
  parallelized by splitting the total population of clones over
  several nodes. The cloning step however creates an overhead since
  one may have to copy clones from one node to another.}.  All in all,
${\cal N}(n,x)$ - the number of clones of in a configuration $x$ at
time $n$ - evolves as
\begin{equation}
{\cal N}(n+1,y) = \sum_x p_\alpha(x,y) k(x){\cal N}(n,x)
\end{equation}

This yields immediately that $Z_t(\alpha)$ is given by
the ratio between the average total population at time $t$
and the population at time $0$ (at initial time every individual 
or clone has type distribution $P(x_0)$)
\begin{equation}
Z_t(\alpha) = \frac{{\cal N}(t)}{{\cal N}(0)} 
\end{equation}
To cope with possible extinction or explosion of the initial population one  
 works with increments \cite{GKP}
\begin{equation}
Z_t(\alpha) =  \frac{{\cal N}(t)}{{\cal N}(t-1)} \frac{{\cal N}(t-1)}{{\cal N}(t-2)} \cdots \frac{{\cal N}(1)}{{\cal N}(0)}
\label{eq:Z_tTfromcloningfactors}
\end{equation}
This allows to keep the population size constant during a simulation
(with a uniform sampling after the cloning with average factor $k(\cdot)$)
and the $Z_t(\alpha)$ will be given by the products of all renormalization
factors. 

{
There are many ways of implementing the Diffusion Monte
  Carlo dynamics described by \eqref{just} and \eqref{justk}, which
  have been extensively discussed in the
  literature~\cite{MJL2007,C2007}. For instance, one may choose to run
  the clones {\em sequentially}, rather than {\em simultaneously}, and
  use any cloning events as the starting point of new
  simulations~\cite{Grassberger}. This makes the algorithm easier to
  parallelize by reducing the overhead but the total number of clones
  is then harder to control.}

\subsection{An example: the totally asymmetric exclusion process}

The Exclusion Process on a lattice consists  of particles which jump to their
neighboring sites at a given rate, conditioned to the fact that the arrival site is 
empty. The large deviations of the total particle currents of a periodic chain of $N$ sites with total asymmetry (TASEP) 
was considered in \cite{GKP}: in this case only jumps to the right are allowed. 

\begin{figure}[ht]
\begin{center}
\includegraphics[width=9.cm]{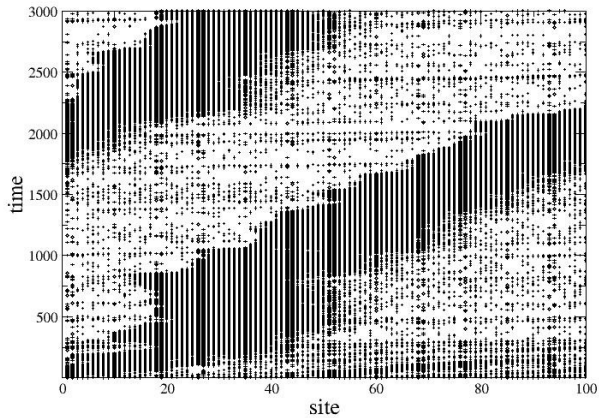}
\end{center}
\caption{{\bf A moving shock in the TASEP}. Space-time diagram for a ring of $N=100$ sites,
$\alpha=-30/N$ and density $0.3$.
The shock drifts to the right.}
\label{uno}
\end{figure}

The technique described above amounts to running various independent
copies of the chain, but cloning a copy in configuration $x$ with an
average rate proportional to
\begin{equation}
k(x)= 1 + \frac{(e^{\alpha}-1)}{N} \times [\mbox{\small number of particles in $x$ with
  a free site to their right}] \label{crist}
  \end{equation}
   The numerical results obtained
for $Z_t(\alpha)$ were compared to the analytic ones of Ref.~\cite{BD}
finding   an excellent agreement with a very modest numerical effort.
Moreover the algorithm allowed to probe the configurations of the
system which are responsible for anomalous small value of the current,
the {\em shocks}, and, in the case of a moving shock,  to follow the
evolution of the second class particle which set the front of the
shock.
\begin{figure}[ht]
\begin{center}
\includegraphics[width=9.cm]{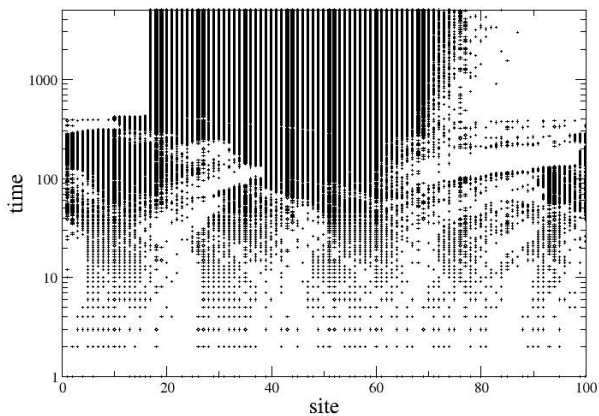}
\end{center}
\caption{{\bf A shock in the TASEP}. Space-time diagram for a ring of $N=100$ sites, $\alpha=-50/N$ and
density $0.5$. Time evolution of a single clone.
The shock is dense and does not advance. Note the logarithmic scale on the $y$-axis.}
\label{cinco}
\end{figure}
In Figure \ref{cinco} we show a space-time diagram of the system
with $N=100$ particles, density $0.5$ and $\alpha=-50/N$. The
simulation was done with $L=1000$ clones, each of them initialized
with random (uniform) occupancy numbers, such that the
configuration had density $0.5$. As predicted by the theory \cite{BD}
for this value of the density, the shock does not drift, although
different initial conditions lead to different shock
positions. Figure \ref{uno} shows the
case $\alpha=-30/N$, and density $0.3$: we see that the shock has a
net drift to the right, again as predicted by the theory.
Let us note here that the configuration corresponds to the end of the
time-interval; but one could have sampled one at an intermediate time
as explained just below Eq (\ref{intermediate}).

The cloning algorithm has been applied for transport models such as
the asymmetric exclusion process and the Kipnis-Marchioro-Presutti
model~\cite{HGprl09,HGjstat09,HGpre10} and to study symmetries in
fluctuations far from equilibrium~\cite{HPPG2010}. Such studies are
useful as a test for the predictions of Fluctuating
Hydrodynamics~\cite{HGprl09,HG2011}, but also to probe the limits of
the cloning method itself, when insufficient clone number may yield
misleading results (a test criterion has been devised in
\cite{HGjstat09}).

\subsection{Continuous-time Markov chains}
\label{subsec:continuous-time-cloning}

Many systems have dynamics that are naturally defined in continuous
time. For instance, spin flips in the Ising model, that takes the
system from a configuration $x$ to another one $y$, can occur at any
time with a given rate $W(x\to y)$. To simulate such systems, one can
discretize  time and the choose a small time step $dt$, (transition
probability writing $p(x,y)=dt W(x\to y)$). One then distinguishes
between time steps during which a configuration change occurs (with
probability, say, $dt W(x\to y)$) and those where nothing happens
(with probability $1-dt\sum_yW(x\to y)$). Doing this in the algorithm
described in the previous sections, one arrives in the limit $dt
\rightarrow 0$ at a continuous time version of the cloning algorithm.

One can however also work directly with continuous time
simulations. Each configuration $x$ has a total escape rate
$r(x)=\sum_y W(x\to y)$, which is the rate at which the system jumps
from configuration $x$ to any other configuration. One can choose a
time interval $\delta t$ from an exponential clock, with probability
$p(\delta t)=r(x) \exp[-r(x) \delta t]$, update the time $t\to
t+\delta t$, and then decide which configuration changes to make. Going
from $x$ to $y$ then occurs with {\it probability} $W(x\to y)/\sum_z
W(x\to z)$. For traditional Monte Carlo algorithms, this method has two
advantages. First, one does not have to decide which $dt$ to
use and the algorithm makes no discretization error. Second, there are no
rejection events  which can slow down severely discrete time
simulations. However all this comes at the cost of having to generate
two random numbers per configuration change (one for the time at which
the change occurs, one for the target configuration) while discrete
time Monte Carlo only needs one.

When simulating rare events, the continuous time method is more
cumbersome to implement but overcomes the problem of diversity of time
scales typically met in these simulations. For instance, depending on
the value of the bias $\alpha$, the TASEP presented above explores
trajectories where the average time between two events ranges from
order $1$ (in a traffic jam, only the leading particle can jump
forward) to order $1/N$ (when all particles can jump forward).  When
working with continuous time, the adjustment of the time-step is automatic. In other
systems, such as the kinetically constrained models presented in
section~\ref{sec:activity}, the situation is even worse. A typical
trajectory can explore successive {configurations} where
the waiting times  may change by a factor
of the order of the system size. In such case, a discrete time algorithm with
a time step small enough to resolve the rapid configuration changes
will have a prohibitively large number of rejection events when visiting
the slow configurations.

To work directly in continuous time, as exposed
in~\cite{LTjstat}, the idea is to write the dynamical partition
function as a sum over allowed values of $F$ (cfr eq (\ref{F})):
\begin{equation}
  Z_t(\alpha)=\langle e^{\alpha F} \rangle = 
  \sum_x \underbrace{\sum_F e^{\alpha F} P(x,F,t)}_{\equiv \hat P(x,\alpha,t)}
  \label{eq:Z_tToflambda}
\end{equation}
where $P(x,F,t)$ is the probability density of being in configuration
$x$ at time $t$, and having observed a value $F$ of the dynamical
observable. The quantity $\hat P(x,\alpha,t)$ is its Laplace
transform.  As in (\ref{questa}), we can choose $F$ to be the sum of
contributions $f(x\to y)$ occurring at each configuration change. For
instance, taking $f(x\to y)=+1$ (resp. $-1$) each time a particle
jumps to the right (resp. left) in a 1d particle system corresponds to
$F$ being the total particle flux flowing through the system from
right to left.  We can also consider the case where $F$ depends on the
time average of some observable $A(x)$, as in the introduction
(see~\cite{LTjstat,LTreview}):
\begin{equation}
  F= \sum_{k=1}^K f\big( x_{k-1}\to x_{k}\big) + \int_0^t dt'\: A(x(t'))
\end{equation}
where $(x_0\ldots x_K)$ is the sequence of visited configurations of
a given history presenting $K$ changes of configurations. $A(x)$ can
for instance be the magnetization of the configuration $x$ of a spin
system and one is then looking for trajectories that have atypical
{\it time average} of the magnetization. 

From the equation of evolution obeyed by $P(x,F,t)$, one obtains
the evolution of~$\hat P(x,\alpha,t)$:
\begin{eqnarray}
  \partial_t\hat P(x,\alpha,t)
  &=& \sum_y e^{\alpha f(y\to x)} W(y\to x) \hat P(y,\alpha,t) 
    \\ && -\sum_y  W(x\to y)\hat P(x,\alpha,t) + \alpha A(x) \hat P(x,\alpha,t)
    \nonumber
\label{eq:evolhatP}
\end{eqnarray}
which is of the form $ \partial_t |\hat P_\alpha \rangle = -  H_\alpha|\hat P_\alpha\rangle $
where $|\hat P_\alpha\rangle$ is the vector of components $\hat P(x,\alpha,t)$.
Just as in Eq (\ref{eq2}), the modified operator of evolution
$H_\alpha$  does not conserve probability  if $\alpha \neq 0$. We have to proceed as
in the steps leading to (\ref{just}) and split the evolution in two
contributions, one conserving probability and the other a purely
cloning term.  To do so we introduce the modified transition
rates $W_\alpha (y\to x)= e^{\alpha f(y\to x)} W(y\to x)$ and the
corresponding escape rate $r_\alpha (x)=\sum_y W_\alpha(x\to y)$. We
can then rewrite~\eqref{eq:evolhatP} as
\begin{eqnarray}
  \partial_t\hat P(x,\alpha,t)
  &=& \overbrace{\sum_y W_\alpha(y\to x) \hat P(y,\alpha,t)
    -r_\alpha(x)  \hat P(x,\alpha,t)}^{\text{probability conserving}}
\nonumber    \\ 
&& + \underbrace{\big[r_\alpha(x)-r(x)+\alpha A(x)\big]  \hat P(x,\alpha,t)}_{\text{cloning}}
\label{eq:evolhatPclone}
\end{eqnarray}

The first part is a modified dynamics of rates $W_\alpha (y\to x)$
while the second part corresponds to cloning at rate
$r_\alpha(x)-r(x)+\alpha A(x)$.  The method is then the same as for
discrete time dynamics (section~\ref{subsec:discretetimecloning}): one
takes a large number of copies of the system, each of them evolving in
continuous time (i) through the modified rates $W_\alpha (y\to x)$ and
(ii) subjected to a cloning probability $e^{[r_\alpha(x)-r(x)+\alpha
    A(x)]\Delta t}$ on each time interval $\Delta t$ where the
configuration does not change from $x$~\cite{LTreview}.  One can
rescale the total clone population to keep its size constant, storing
as previously  the overall cloning factor.  The dynamical partition
function is then recovered from those factors as
in~\eqref{eq:Z_tTfromcloningfactors} and the corresponding dynamical
free energy $\mu(\alpha)$ is:
\begin{equation}
  \mu(\alpha)=\lim_{t \to\infty} \frac 1t \log Z_t(\alpha)
\end{equation}
 We provide in Appendix~\ref{appendix-pseudo-code} an example pseudo-code
for the practical implementation of the algorithm.

\subsection{An example: density profiles in the ASEP}

\begin{figure}[ht]
\begin{center}
\includegraphics[width=.6\columnwidth]{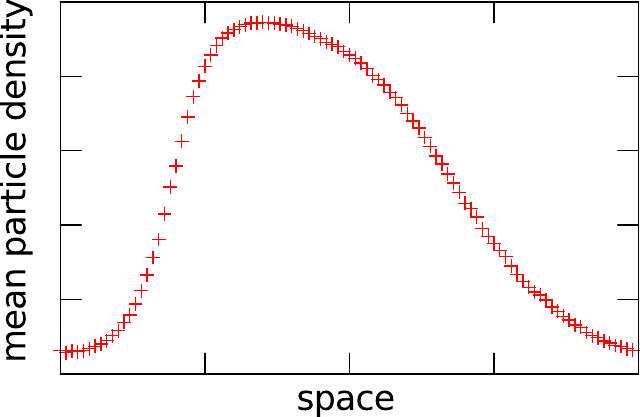}
\end{center}
\caption{{\bf Density profile in the ASEP}. 
$\alpha=-0.3$, $\alpha$ conjugated
to the total current flowing through the system. System size is 400, with 200 particles, in
periodic boundary conditions, with an asymmetry $E=\frac 12 \log \frac pq = -0.2.$}
\label{fig:asepprofile}
\end{figure}

Exclusion processes (such as the TASEP studied above) are interesting
transport models in which the cloning algorithms can be used and in
particular compared to analytical results for the cumulant generating
function $\mu(\alpha)=\lim_{t \rightarrow \infty} \ln
Z_t(\alpha)/t$~\cite{LTjstat,HGprl09}, including finite size
effects~\cite{LTreview}.  In Fig.~\ref{fig:asepprofile}, we present an
example of a mean profile at non-zero $\alpha$ for the asymmetric
exclusion process (compared to the TASEP, particles can jump to the
left and to the right with respective rates $p$ and $q$). The
parameter $\alpha$ is conjugated to the particle flux through the
system.  We observe on Fig.~\ref{fig:asepprofile} that, to minimize
the overall current, the system develops an asymmetric profile, where
only the front particles can jump easily.



\section{Fluctuations of Dynamical Activity}
\label{sec:activity}



Driven systems  may reach a non-equilibrium \emph{steady
  state}, characterized by a non-zero current  the probability distribution 
of which  can be studied as described in the previous section.
Another class of non-equilibrium systems is given by
glassy systems. In the most simple cases, these systems are out of
equilibrium not because they are driven but because their dynamics is so
slow that a macroscopic system never reaches Boltzmann equilibrium (or any other steady state),
despite the fact that the microscopic dynamics satisfy detailed balance.  In this context, it can be interesting to study
trajectories of atypical mobility, for instance to detect trajectories
 that are `faster' or 'slower' than average, i.e. the dynamic heterogeneity. To quantify this, one introduces 
 the \emph{dynamical activity}~\cite{MGC2005, LAvWcras2007, BL2008}
(also termed traffic~\cite{MNW2008,MN2008}), which provides a good
description of dynamical heterogeneity in glass models, as we now
discuss.

On a time window $[0,t]$ the dynamical activity $K$ of a stochastic
process is the number of
configuration changes undergone by the system, and is thus a random
variable that depends on the system's trajectory.

Kinetically constrained models (KCMs), such as the
Fredrickson-Andersen~\cite{FA1984} or the Kob-Andersen\cite{KA1993} models
are such that static (one-time) properties are
trivial in the most simple cases,  while their dynamical properties
(\emph{e.g.} two- or more times correlations) share common features
with generic glassy phenomena
(see~\cite{RS_reviewKCM2003,GST_reviewKCM2011} for reviews on KCMs).
They lend themselves rather easily for the study of their 
  activity  $K$, and for the analysis of the results.  
\begin{figure}[h]
\begin{center}
\includegraphics[width=.5\textwidth]{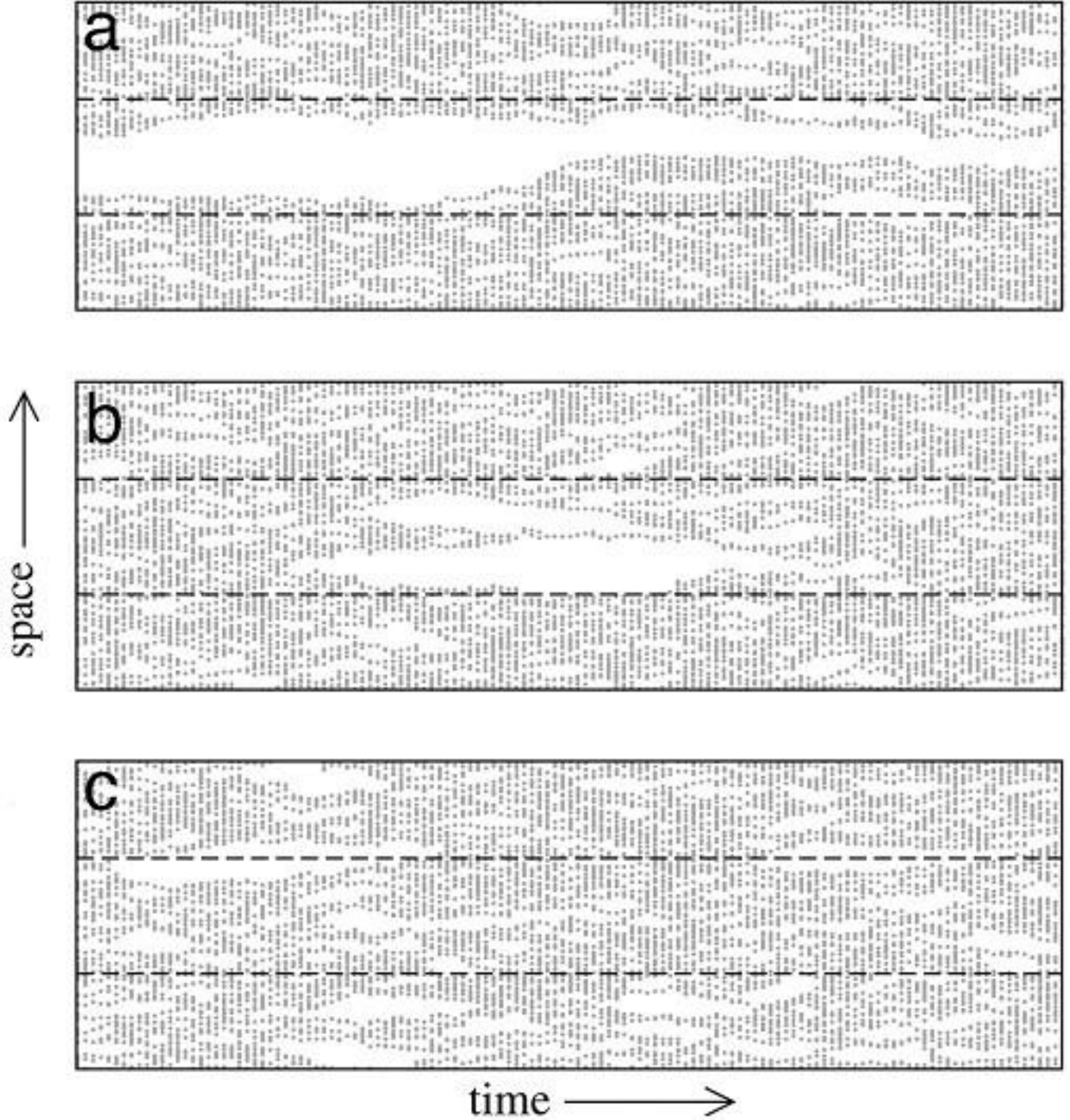}
\end{center}
\caption{{\bf FA model (From Merolle \emph{et al.}~\cite{MGC2005})}.
Space-time diagram of the FA model for atypical (a and b) and
typical (c) histories. In the space direction, active sites
are represented in black dots while inactive ones are white.
The picture is reminiscent of the phase coexistence of
a static medium at a solid-liquid coexistence point.
}
\label{fig:spacetime-bubbles}
\end{figure}

Let us focus for simplicity on the one-dimensional Fredrickson-Andersen
(FA) model.  It consists in a 1d lattice of $L$ sites. Each site is
either excited  (low density, active)  or unexcited  (high density, inactive).  The sites may flip from inactive to active (at rate $c$),  and 
from active to inactive (at rate $1-c$).
 These transitions are allowed on a given
site provided at least one of the neighboring sites is active. 
This
is the \emph{kinetic constraint}, introduced as a way to mimic  the facilitated
dynamics of molecular glasses, whereby  active regions enhance activity in their
neighborhood.  Clearly, for small values of $c$, the dynamics becomes very slow.

 It was observed in~\cite{MGC2005} that the FA model
presents ``dynamical coexistence'' of active and inactive regions in
space-time (see Fig.~\ref{fig:spacetime-bubbles}), very similar
to the phase coexistence of liquid and solid at the coexistence 
point in a first order static phase transition --~ if one forgets that one
direction in Fig.~\ref{fig:spacetime-bubbles} is the time.

The  activity $K$ of a configuration is defined as the number of active sites.
In practice, one may weight the
{trajectories} followed by the system by a factor $e^{-s K}$, to
favor active ($s<0$) or inactive ($s>0$) histories (in this section we take the
convention $s=-\alpha$ to follow  the notation in the  literature on
KCMs).  If the observed
coexistence disappears for $s\neq 0$ (that is, if there is a dynamical phase
transition), it means that the system indeed sits on a first-order
dynamical coexistence point at $s=0$.

The continuous time cloning algorithm~\cite{LTjstat} exposed in
section~\ref{subsec:continuous-time-cloning} enables us to compute numerically the dynamical partition function
\begin{equation}
  Z_t(s)=\langle e^{-sK}\rangle\sim e^{t \mu_L(s)}
\end{equation}
for this system and other
KCMs~\cite{GJLPvDvW2007,GJLPvDvW2009} 
The average is taken on histories of duration $t$, in the large $t$
limit, at fixed system size $L$.  The non-analyticities of
the dynamical free energy $\mu_L(s)$  in the large-size limit,  signal  the
existence of a  
dynamical phase transition.

\begin{figure}[ht]
\begin{center}
\includegraphics[width=\columnwidth]{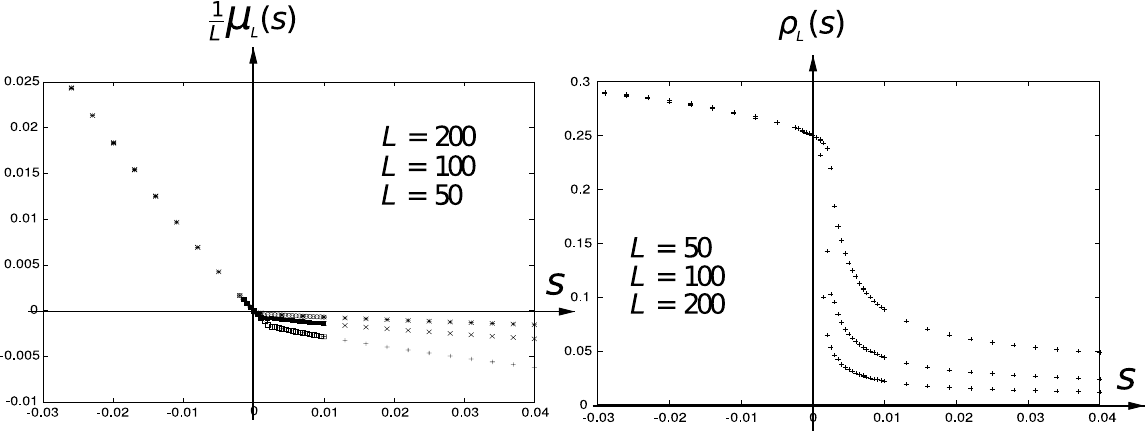}
\end{center}
\caption{{\bf FA model (From Garrahan \emph{et al.}~\cite{GJLPvDvW2007})}. 
Numerical
  evaluation of the ``dynamical free energy'' $\frac 1L\mu_L(s)$
  (left) and the density of occupied sites $\rho_L(s)$ (right) for histories
  weighted by $e^{-sK}$. As the system size increases, $\frac
  1L\mu_L(s)$ approaches its singular limit where $\frac 1L\mu_L(s)$
  is zero for $s>0$. In the same limit, the density $\rho_L(s)$
  displays a discontinuity at $s=0$, indicating a first order dynamical
  transition.  }
\label{fig:numerics_FA}
\end{figure}

\subsection{Dynamical phase coexistence }

As shown in~\cite{GJLPvDvW2007,GJLPvDvW2009}, several KCMs display a
phase transition, in the large system size limit, between an active
phase ($s\leq 0$) where the dynamical free energy $\frac
1L\mu_L(s)$ is finite, and an inactive phase ($s>0$) where is
identically zero (see Fig.~\ref{fig:numerics_FA}, left, for the 1d FA model).
The mean density of active sites 
%
(see Appendix~\ref{appendix-pseudo-code} for details
on the practical computation of such a weighted mean)
%
\begin{equation}
  \rho_L(s)= \frac{\langle e^{-sK}\frac 1t \int_0^t \frac 1L\sum_{i=1}^Ln_i\rangle}{\langle e^{-sK}\rangle}
\end{equation}
(here $n_i\in\{0,1\}$ is the activity  at site $i$) also
characterizes this transition (Fig.~\ref{fig:numerics_FA}, right): it
remains finite in the active phase $s\leq 0$ (a finite fraction of
sites is active) while it goes to zero in the inactive phase (only a
finite number of sites remains inactive).  Several other glass formers
display the same phenomenology (see~\cite{CGannrev} for a review),
representative of dynamical heterogeneities, that is, of the
coexistence in the system of regions with high and low dynamical activity.

An interesting question is to determine whether molecular models of
glasses, such as Lennard Jones mixtures, also present such a dynamical
phase transition.  A conceptual difficulty that arises  is
to find a physically relevant measure of the mobility, that generalizes  the
concept of dynamical activity to this context. In~\cite{HJGC}, the
activity was defined as the number of  events where particles move sufficiently far in a given time-interval,
 thus averaging out short-scale vibrations, whereas in
\cite{PLvW}, the activity was taken to be  a  time-average of the modulus of the forces, in a continuous version of the model.
In both approaches, numerical results support the existence of a 
phase transition at some critical value $s_c$.
An open issue is to characterize the inactive phase and to determine
whether the effective finite-size critical transition parameter
$s_c(L)$ goes to $0$ as $L$ goes to infinity or not (that is to say:
does the standard dynamics at $s=0$ lie exactly at the critical point?).
%

More generically, the phase transitions are also present in $p$-spin
models~\cite{vDJvW} and in trap models~\cite{vDSvW}, where numerical
approach support analytical results. These results are in favor of a
generic link between glassiness and dynamical phase coexistence, whose
precise nature remains to be understood.
%
%
%

\section{Fluctuation of chaoticity in dynamical systems}
\label{chaos}

As explained in the previous section, large deviation theory plays
nowadays an important role in non-equilibrium statistical physics
to study and quantify dynamical phase transitions. The first studies
of large deviations of dynamical observables were however inspired by
another field, that of dynamical systems. It was argued in the 70s,
following the seminal works of Sinai, Ruelle, Bowen and
others~\cite{Sinai72,Bowen75,Ruelle76,Ruelle78} that quantitative
studies of dynamical systems should rely on a construction analogous to
statistical mechanics of trajectory space, where  the quantities playing the  role of energy functionals
for the trajectories are
 functions of  the Lyapunov exponents.  This line of
thought was very successful in terms of formalism and theory, but
progress was severely hampered by the difficulty of computing anything 
in all but the most schematic systems. Indeed, many of the examples
studied  very low dimensional
systems ---  mostly  maps of the interval,  with notable exception of the Lorenz
gas~\cite{Beijeren2005}. As we show in the two following sections, the
development of recent methods to compute the fluctuations of Lyapunov
exponents can fill this gap and hopefully lead to new insights  in
the field of dynamical systems of {\em many} bodies.

For sake of concreteness, we will focus on Hamiltonian dynamics but
one should keep in mind that the method is much more general and can
be applied, for instance, to dissipative systems. We consider a system
with $2N$ degrees of freedom whose dynamics is given by
\begin{equation}
  \dot x_i = f_i[{\bf x}(t)];\quad\mbox{with}
  \begin{cases}
    {\bf x}=(q_1,\dots,q_N,p_1,\dots,p_N)&\\
    {\bf f}=(\frac{\partial H}{\partial p_1},\dots,\frac{\partial H}{\partial p_n},-\frac{\partial H}{\partial {q_1}},\dots,-\frac{\partial H}{\partial {q_N}})&
   \end{cases}
\end{equation}
As usual to quantify the chaoticity of a trajectory we introduce the
Lyapunov exponents. We consider an infinitesimal perturbation
${\bf \delta x}(t)$ whose dynamics reads
\begin{equation}
  \delta \dot {\bf x} = -A \cdot \delta \bol x;\quad\text{with}\quad
  A_{ij}=-\frac{\partial f_i[\bol x(t)]}{\partial x_j}
\end{equation}

The evolution of the norm of such a perturbation is given by
\begin{equation}
  \label{eqn:evolnorm2}
  \frac{\d}{\dt} |\delta {\bf x}|^2= -\sum_{ij} 2 \,\delta x_i\, A_{ij}\,
  \delta x_j
\end{equation}
Introducing the normalized tangent vectors $v_i=\frac{\delta
  x_i}{|\delta {\bf x}|}$ whose evolutions are given by
\begin{equation}
  \label{eqn:evolvecttangnorm}
  \dot v_i = - \sum_j A_{ij}v_j + v_i \sum_{kl} v_k A_{kl} v_l
\end{equation}
equation \eqref{eqn:evolnorm2} can be recast as
\begin{equation}
  \frac{\d}{\dt} |\delta {\bf x} (t)|^2  = -\sum_{ij} 2 v_i A_{ij} v_j
  |\delta {\bf x}(t)|^2
\end{equation}
and finally solved to yield
\begin{equation}
  | \delta {\bf x}(t)|=|\bol \delta {\bf x}(0)|
  \ee^{-\sum_{ij}\int_0^t v_i(t') A_{ij}[\bol x(t')] v_j(t') \dt'}
\end{equation}
The largest Lyapunov exponent is then given by
$\lambda = \underset{t\to\infty} \lim \lambda(t)$,
 where the finite time Lyapunov exponent $\lambda(t)$ is
 \begin{equation}
  \label{eqn:deflambda}
\lambda(t)=\frac 1 t \log\frac{|\delta {\bf
      x}(t)|}{|\delta {\bf x}(0)|}
  =- \frac 1 t \int_0^t \dt' \Big\{ \sum_{ij} v_i(t') A_{ij}[\bol x(t')] v_j(t')\Big\}
\end{equation}
More generally, the exponential expansion of  $k$-dimensional volume elements, rather
that vectors $\delta {\bf x}$, yields in a similar way the sum of the first $k$
 Lyapunov exponents.

To characterize the fluctuations of chaoticity amounts to sampling the
distribution of $\lambda(t)$
\begin{equation}
  \label{eqn:ldfpi}
  P(\lambda,t) = \ee^{S(\lambda,t)}\underset{t\to\infty}\sim \ee^{t s(\lambda)}
\end{equation}
One can understand that the exponent is
generically extensive in time, as in usual thermodynamic systems: 
one cuts a long trajectory of
duration $t$ in many segments of duration $\delta t$ much larger than the 
typical correlation time $\tau$. Each segment can thus be considered independent of the others
and the probability that the total trajectory has an exponent
 $\lambda$ is
 \begin{eqnarray}
  P(\lambda,t)
  &=&\!\!\!\!\!\!\!\!\!\sum_{(\lambda_1+\dots+\lambda_{t/\delta t})\delta t=\lambda
    t}\!\!\!\!\!\!\!\!\! P_1(\lambda_1,\delta t)\dots
  P_{t/\delta t}(\lambda_{t/\delta t},\delta t)\\
&=&\!\!\!\!\!\!\!\!\!\sum_{(\lambda_1+\dots+\lambda_{t/\delta t})\delta t=\lambda t}\!\!\!\!\!\!\!\!\! \ee^{S_1(\lambda_1,\delta t)+\dots +S_{t/\delta t}(\lambda_{t/\delta t},\delta t)}
\end{eqnarray}
The exponent of each term of the r.h.s. is the sum of $t/\delta t$
terms of order one and is thus of order $t$. At large times, $t/\delta t\gg 1$, the distribution 
$P(\lambda,t)$ concentrates around its typical value,  and the scaling 
law (\ref{eqn:ldfpi}) is thus verified. This scaling breaks down in the presence of diverging correlation
times, a signature of dynamical phase transitions.

As in statistical mechanics, the derivation of the entropy $s(\lambda)$ is
difficult and one rather works in a ``canonical'' ensemble by
introducing a dynamical partition function
\begin{equation}
  \label{eqn:Z_t}
  Z_t(\alpha) = \left\langle \ee^{\alpha t \lambda(t)} \right \rangle\underset{t\to\infty}\sim \ee^{t \mu(\alpha)}
\end{equation}
where the average $\langle\, .\, \rangle$ is made with respect to
$P(\lambda,t)$, i.e. over initial conditions, noise realizations,
etc. $\mu(\alpha)$ plays the role of $-\beta F$ in statistical
mechanics, where $F$ is a free energy, and is called topological
pressure.

From the definition of the finite time Lyapunov exponent
\eqref{eqn:deflambda}, one sees that the computation of $Z_t(\alpha)$
amounts to the large deviation computation presented in the
introduction, with the observable ${A}$  now given by
\begin{equation}
  {A}({\bf x})=-\sum_{i,j} v_i A_{ij}({\bf x}) v_j;\qquad F=\int dt A({\bf x})
\end{equation}

Let us now make a point that will be valid for all deterministic systems. In such  cases, the only source of fluctuations are the initial conditions. If the system is chaotic enough, this should not be very important but, for example, in the case of mixed system, starting from a regular island or a chaotic region yields a very different result, because trajectories do not take from one to the other. In this review we consider a shortcut to this problem which consists of adding a small amount of stochastic noise,  so that the dynamics effectively samples the whole trajectory space
(for a discussion of the low noise limit see
\cite{K07}). 
We thus consider a slightly different set of equations
\begin{equation}
\label{eqn:hamequabruit}
  \dot q_i=p_i;\qquad \dot p_i = -\frac{\partial H}{\partial q_i} + \sqrt{2 \eps} \eta_i
\end{equation}


The algorithm presented in the introduction of this paper can now be
applied to our noisy Hamiltonian dynamics. We consider a population of
${\cal N}$ clones in phase space of positions and momenta
$\bol q$ and $\bol p$. To each clone we associate a normalized tangent
vector $\bol v$. We then choose a time step $\dt$ and a noise
intensity $\epsilon$ and run the simulation over a large time
$t=M\dt$. At $t=0$, the ${\cal N}$ copies of the system start from an arbitrary
initial configuration (the noise ensures the ergodicity of the algorithm).
At each time step $t'=n \dt$, we do the following~\cite{TK2007}:
\begin{itemize}
\item[\fbox{$1$}] For each clone
  \begin{itemize}
  \item[$\bullet$]  $(\bol q, \bol p)$ evolve with the noisy Hamiltonian dynamics (\ref{eqn:hamequabruit}),
  \item[$\bullet$] $\bol v$ evolves according to the
    linearized dynamics
    \begin{equation}
      \dot v_i = -A_{ij} v_j
    \end{equation} 

  \item[$\bullet$] $\bol v$ is then renormalized to unity and we store
    the renormalization factor $N(n)=\frac{|{\bf v}(t+dt)|}{|{\bf v}(t)|}\simeq \ee^{-{\bf v}^\dagger \cdot A \cdot {\bf v} \dt}$.
  \end{itemize}
\item[\fbox{$2$}] Each clone of the system is then pruned or replicated, with its
  rate $N(n)^\alpha$. To do so, we pull a random number $\epsilon$
  uniformly between 0 and 1 and we compute\footnote{$\lfloor x
    \rfloor$ is the largest integer smaller than $x$} $\tau=\lfloor
  \eps+N(n)^\alpha \rfloor$,
  \begin{itemize}
  \item[$\bullet$] if $\tau=0$, the clone is deleted
  \item[$\bullet$] if $\tau>1$, we create $\tau-1$ copies of the clone
  \end{itemize}
\item[\fbox{$3$}]  The total population is now composed of ${\cal
  N}(n+1)$ clones, instead of the initial ${\cal N}(n)$ ones. We then store $R(n)=\frac{{\cal N}(n+1)}{{\cal N}(n)}$,
  \begin{itemize}
  \item[$\bullet$] if ${\cal N}(n+1)<{\cal N}(n)$, we copy
     ${\cal N}(n+1)- {\cal N}(n)$ clones, chosen at random,
  \item[$\bullet$] if ${\cal N}(n+1)>{\cal N}(n)$, we delete
     ${\cal N}(n+1)-{\cal N}(n)$ clones, chosen at random,
  \end{itemize}
Finally, we end up again with ${\cal N}(n+1)={\cal N}(n)={\cal N}(0)$ clones.
\end{itemize}
The dynamical partition function is then obtained from $R(n)$ through
\begin{equation}
  Z_t(\alpha)=\prod_{n=1}^M R(n)
\end{equation}
while the topological pressure is given by
\begin{equation}
  \mu_t(\alpha)= \frac 1 t \sum_{n=1}^M \log R(n)
\end{equation}
Let us now illustrate this algorithm, called ``Lyapunov Weighted
Dynamics'', with a low dimensional system (the standard map) and a large
dimensional one (a FPU chain of 1024 particles).
\begin{figure}[ht]
  \begin{center}
    \begin{minipage}{.49\textwidth}
      \includegraphics[width=\textwidth]{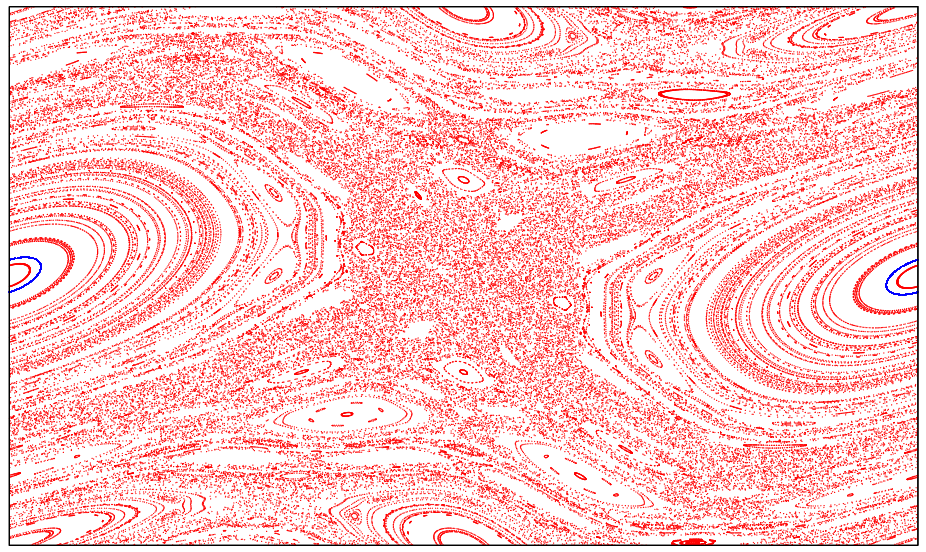}
    \end{minipage}
    \begin{minipage}{.49\textwidth}
      \includegraphics[width=\textwidth]{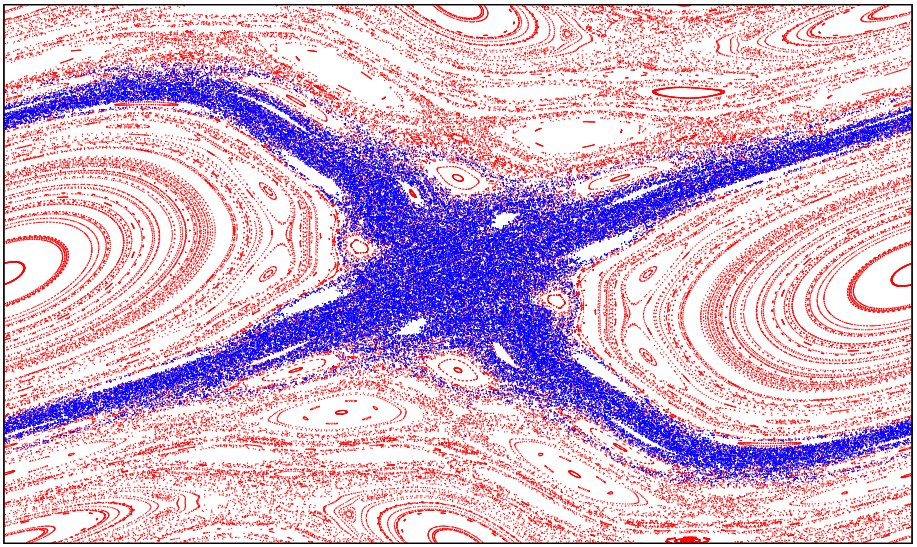}
    \end{minipage}
    \begin{minipage}{.49\textwidth}
      \centering $\alpha=-0,04$
    \end{minipage}
    \begin{minipage}{.49\textwidth}
      \centering $\alpha=0,04$
    \end{minipage}
    \caption{{\bf Typical Configurations for $\alpha= \pm
        0,04$}. Phase space trajectories of the standard map are shown
      in light red whereas the trajectories localized by the Lyapunov
      Weighted Dynamics appear in dark blue.}
    \label{fig:standardmap1}
  \end{center}
\end{figure}

\subsection{The Standard Map}
The standard map is defined by the dynamics
\begin{equation}
  p_{n+1}=p_n+\frac{k \delta}{2\pi} \sin(2 \pi q_n);\qquad q_{n+1}=q_n+\delta p_{n+1}
\end{equation}
with $(q_n,p_n)\in [0,1]\times[-1,1]$. It is one of the traditional models
used to study transition to chaos. It goes from an integrable system
when $k=0$ to a more and more chaotic one when $k$ increases. In
figure~\ref{fig:standardmap1} we show the typical trajectories that
are localized by the Lyapunov Weighted Dynamics for very small bias
($\alpha=\pm0.04$). One sees that as soon as the system is biased in
favor of integrable trajectories ($\alpha<0$), the dynamics localizes on
integrable islands, whereas a tiny bias favoring chaotic
trajectories ($\alpha>0$) detects the chaotic layers surrounding these
islands.

\begin{figure}[ht]
  \begin{center} 
    \includegraphics[width=.8\textwidth]{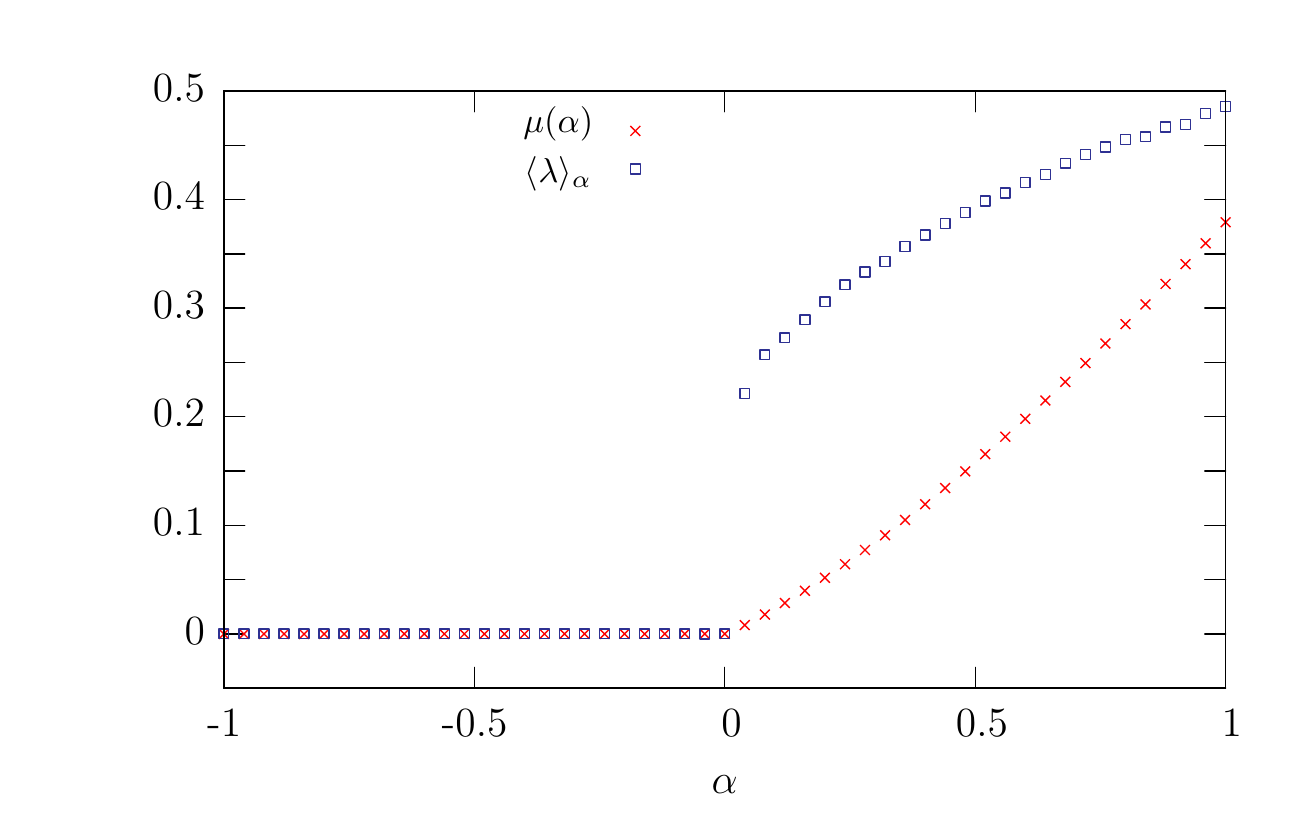}
  \caption {{\bf Standard map}. Dynamical free energy $\mu(\alpha)$ (red crosses) and
    average Lyapunov exponent $\langle \lambda
    \rangle_\alpha=Z_t^{-1}(\alpha) \langle \lambda \ee^{\alpha \lambda t}\rangle$ 
    (blue squares) as a function of the bias $\alpha$. The
    discontinuity at $\alpha=0$ of $\langle \lambda
    \rangle_\alpha=\mu'(\alpha)$ is the signature of a phase coexistence
    between chaotic and integrable trajectories in space time.}
    \label{fig:standardmap2}
  \end{center}
\end{figure}

Computing the topological pressure (figure~\ref{fig:standardmap2})
shows that the system lies at a critical point where chaotic and
integrable trajectories coexist in phase space, in the manner of a
first order phase transition.

\subsection{FPU chains}

\begin{figure}[ht]
  \includegraphics[width=\textwidth]{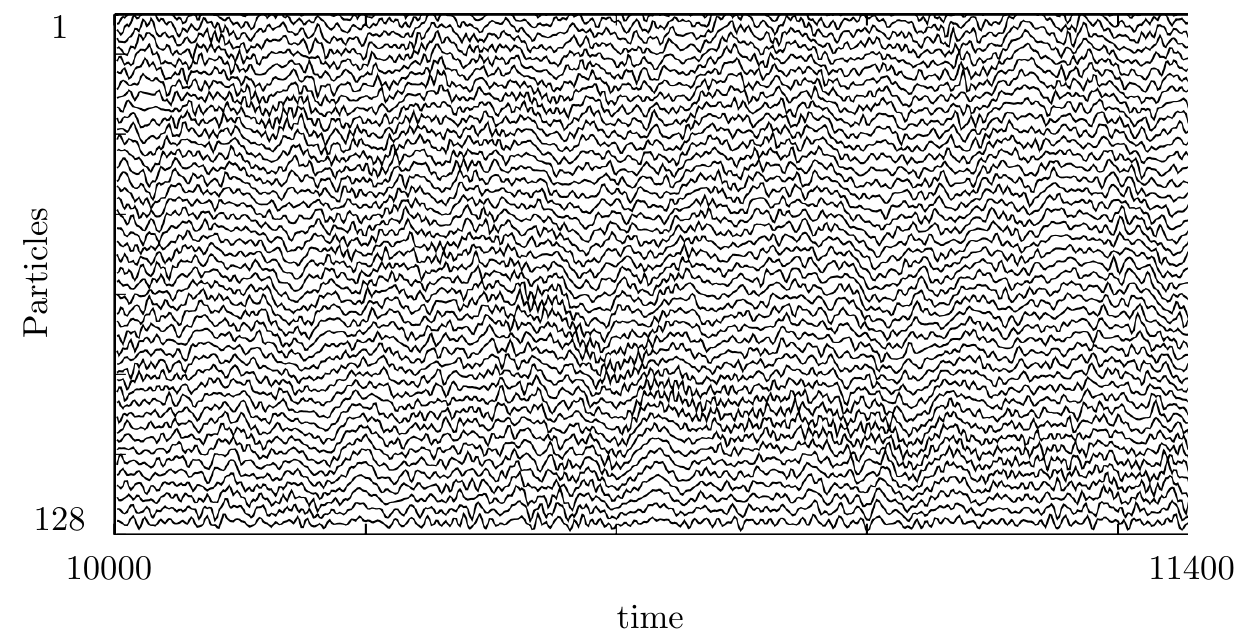}
  \begin{center}
    \caption{{\bf Equilibrium simulations of the FPU chain ($N=128,\,
        \alpha=0$)}. Time-line of each of the 128 particles around
      their [arbitrary] equilibrium positions. We see a superposition
      of localized breathers, ballistic solitons and small
      fluctuations.}
    \label{fig:FPUequilibre}
  \end{center}
\end{figure}

Beyond the computation of dynamical free energies (or topological
pressure), the algorithm can be used to sample trajectories of
atypical chaoticity. Let us show here on a high-dimensional system,
with 2048 degrees of freedom, which are the trajectories that realize
large deviations of the chaoticity in anharmonic chains of
oscillators. We consider the following Hamiltonian
\begin{equation}
  \H= \sum_{i=1}^N \frac {p_i^2}{2} + \sum_{i=1}^{N} 
  \left[\frac{(x_{i+1}-x_i)^2}2 + \beta \frac{(x_{i+1}-x_i)^4}4\right]
\end{equation}
where $x_{N+1}=x_1$. This system, studied in the 50s by Fermi, Pasta,
Tsingou and Ulam, corresponds to $N$ particles connected by anharmonic
springs. The limit $\beta=0$ corresponds to an integrable case: the
springs are harmonic and the Fourier modes correspond to $N$
independent harmonic oscillators or frequencies
\begin{equation}
  \omega_k = 2 \sin \left(\frac {\pi k}{N}\right)
\end{equation}
There has been  continuous interest in this model  (for a
review see~\cite{Berman05}) because of its rich phenomenology,  and
in particular, there has been some recent studies of the (Gaussian) fluctuations of
its Lyapunov exponent \cite{Politi2011}. As soon as $\beta$ is non-zero, the
dynamics are chaotic. However, starting from well chosen initial
conditions, the model admits long-lived solitonic modes, related to
the Korteweg-de~Vries modified
equation~\cite{Kruskal64}. Similarly, a modulational
instability leads to short-lived chaotic
breathers~\cite{Cretegny98,Trombettoni01}, when energy is injected in
high-frequency modes. If one runs an equilibrium simulation of the
anharmonic chain, one typically observes a mixture of short-lived
localized structures (solitons, breathers) and a phonon bath (figure
\ref{fig:FPUequilibre}).

When applying the Lyapunov Weighted Dynamics, we  add
a small stochastic noise to the system, taking care that the  noise conserves the total
energy and momentum and thus  preventing  a slow, unphysical drift in these quantities.

\begin{figure}[ht]
  \begin{center}
    \includegraphics[width=.9\textwidth]{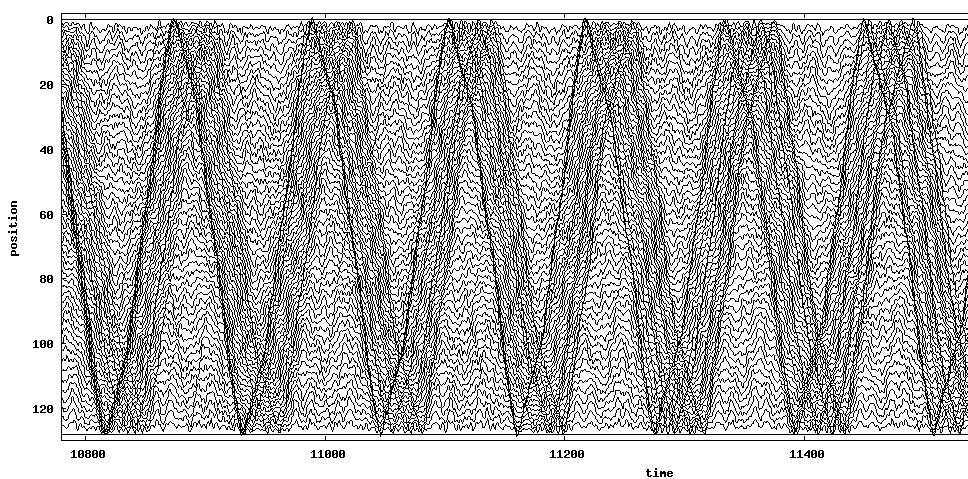}
    \caption{ {\bf Looking for regular trajectories ($N=128,\,
        \alpha=5\,N$).}  Simulation at fixed energy ($E=1$) with fixed
      boundary conditions, starting from microcanonical
      equilibrium. The figure shows the time-line of each particles
      around its [arbitrary] equilibrium position. Several solitons
      are ballistically propagating from one end of the system to the
      other, where it elastically bounce of fixed boundary
      condition. the Lyapunov exponent of this trajectory is equal to
      half the average one.}
    \label{FPUregular}
  \end{center}
\end{figure}

If one biases the system in favor of regular trajectories, the phonons
and breathers completely disappear and we observe a long-lived gas of
solitons, propagating ballistically (see figure~\ref{FPUregular}). In this case, it is important to set the center of mass velocity to zero, because otherwise the system can eliminate completely chaoticity by concentrating all its energy on the center of mass motion.

\begin{figure}[ht]
  \begin{center}
    \includegraphics[width=\textwidth]{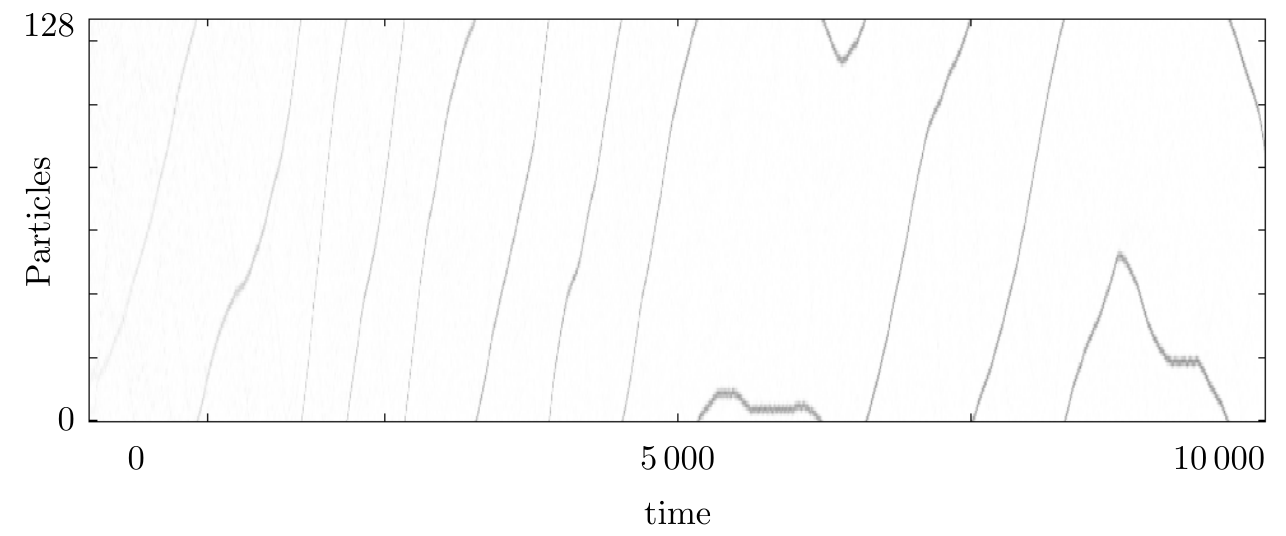}
    \caption{{\bf Looking for chaotic trajectories ($N=128,\,
        \alpha=5\,N$)} This simulation is realized at fixed energy and
      total momentum (energy density $E=1$) with periodic boundary
      conditions. The gray level represent the total energy of each
      particles. Starting from an equilibrium configuration, the
      dynamics reveals a chaotic breathers whose Lyapunov exponent is
      three time larger than the average one.}
    \label{FPUbreatherN128}
  \end{center}
\end{figure}

\begin{figure}[ht]
  \begin{center}
    \includegraphics{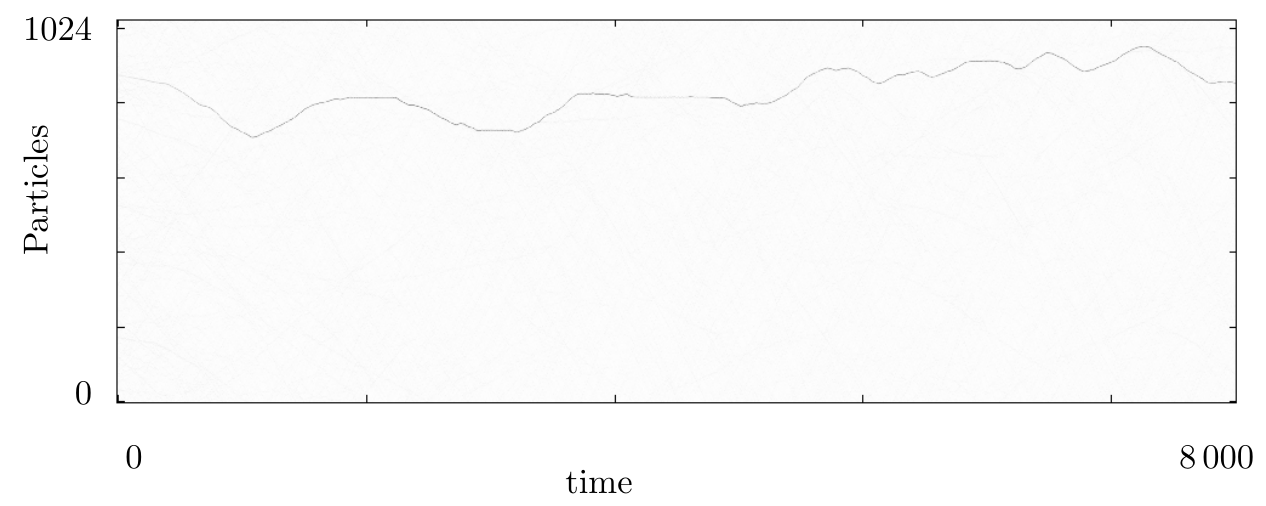}
    \caption{{\bf Looking for chaotic trajectories ($N=1024,\,
        \alpha=5\,N$) }. This simulation is realized at fixed energy and
      total momentum (energy density $E=1$) with periodic boundary
      conditions. The gray level represent the total energy of each
      particles.}
    \label{FPUbreatherN1024}
  \end{center}
\end{figure}

On the other extreme, a bias in favor of chaotic trajectories
localizes long-lived chaotic breathers (see figure
\ref{FPUbreatherN128}). {
We used periodic boundary
  conditions for this simulation to reduce the interactions between
  the wandering breather and the boundaries of the system}. Note that
  running the same simulation in a much larger system (N=1024) shows
  that the breathers are much more localized than the solitons (figure
  \ref{FPUbreatherN1024}).

Interestingly, the values of the bias $\alpha$ we have to use here are not of order
one. Indeed, as $N$ increases, the distribution of the largest
Lyapunov exponent becomes more and more peaked. Let us assume for
instance that $s(\lambda)$ is extensive with some power of the system
size, so that one can write
\begin{equation}
  P(\lambda_1,t) = \exp[N^\xi t \tilde s(\lambda_1)]
\end{equation}
with $\tilde s(\lambda)$ of order 1 in both $t$ and $N$. From the expression
\begin{equation}
  Z_t(\alpha) = \left \langle \ee^{\alpha \lambda t} \right \rangle = \int d\lambda \exp[N^\xi t \tilde s(\lambda_1)+\alpha \lambda t]
\end{equation}
one sees that the integral is dominated by a value $\lambda^*$ such that:
\begin{equation}
  \tilde s'(\lambda^*) = - \frac{\alpha}{N^\xi}
\end{equation}
When $N\to \infty$, $\lambda^*$ satisfies $s'(\lambda^*)=0$ and
is thus the typical value of the Lyapunov exponent. One should thus use
a bias that scales as $\alpha= N^\xi \tilde \alpha$ to observe large
deviations of the Lyapunov exponents. Similarly, to access the
dynamical free energy, one has to compute the exponent $\xi$ and
define
\begin{equation}
  \tilde \mu(\tilde \alpha)=\frac 1 {t N^\xi} \log Z_t(\alpha)
\end{equation}
Such a calculation, which, as far as we know, has not been done so
far, would tell if the FPU chain lies at a critical point where
breathers, solitons and phonons coexist in a first order phase
transition manner. The computation of the dynamical free energy for
large dimensional systems is now achievable numerically and is one of
the exciting goal that are facing us.

\section{Work and entropy production}
\label{gallavotti-cohen}

When a system is subjected to an external drive, the total energy
absorbed (and the resulting entropy production), are quantities that
fluctuate depending on the initial microscopic configuration of the
system and on the thermal bath, if there is one.  Work and entropy
production are important quantities, because they concern the state of
the system and are the subject of the Second Law of thermodynamics.
The Second Law as such concerns only average quantities, and not the
fluctuations. It was only relatively recently realized that a wider
framework -- based on considering the effect of time-reversal on the
dynamics -- allows to derive a set of relations that are obeyed by the
fluctuations -- well beyond the linear regime -- and yields the Second
Law constraints as particular cases.

{\em i)}  The {\em transient}  Fluctuation Theorem relates, in the same context, the probability of a given work $W$, and that of its opposite:
$P(W)/P(-W) = e^{W/T}$ \cite{ECM,GC}. 

{\em ii)} The Jarzynski relation states that the average of $e^{- W/T}$   over all processes starting from an equilibrium distribution at temperature $T$ is {\em one} \cite{Jarzynski}.

Both are very general, model-independent results, and were  later shown to be  particular cases of the more general relation,  Crooks'  relation.

{\em iii) }The {\em stationary} fluctuation theorem involves the same relation for the work as the transient version,  in a stationary  (non-equilibrium) situation,   and is valid only in the limit of large times.
The particular case in which the dynamics is deterministic (the Gallavotti-Cohen theorem \cite{GC}) 
deserves special attention: the theorem   is non trivial because
the nature of the stationary distribution is then dependent upon the ergodicity properties of the system.
These conditions involve not only chaoticity properties of the attractor, as one would expect from any  problem  in ergodic theory,
but also the fact that attractor and repellor sets are sufficiently intertwined: large deviation trajectories  that commute between    
them generate the reversals in entropy production \cite{JK}.

Systems with macroscopic, {\em hydrodynamic} degrees of freedom  may have extremely large fluctuations when subjected to strong forcing, due to excitation of macroscopic structures \cite{pinton}. 
The typical example is the (Rayleigh-B\'enard) convection of a fluid between a hot $T_h$ lower plate and a colder $T_c$ top plate \cite{sergio}. The heat is transported by  fluid currents that have macroscopic fluctuations, enormous compared  with  $k_b T_h$.
The fluctuation theorem as such involves the temperatures $T_h,T_c$ that are   irrelevant for these fluctuations. The only way in which
 the appearance of a Fluctuation Relation for the hydrodynamic modes may be justified, is to invoke the existence of a large  {\em effective temperature}, related to
the macroscopic fluctuations.
Bonetto and Gallavotti  \cite{BG} have conjectured that this could be justified by considering the restricted space in which the
macroscopic  takes place.
These questions are very much open, and in order to make progress it would be useful  to simulate
the  limits beyond which the fluctuation theorem ceases to  hold rigorously, because 
that is where new concepts may arise. These are the limits in which large deviations are particularly hard to observe, if one has to wait for them to happen spontaneously.

\subsection{Sinai billiard}

\begin{figure}[ht]
\begin{center}
\includegraphics[width=6.cm]{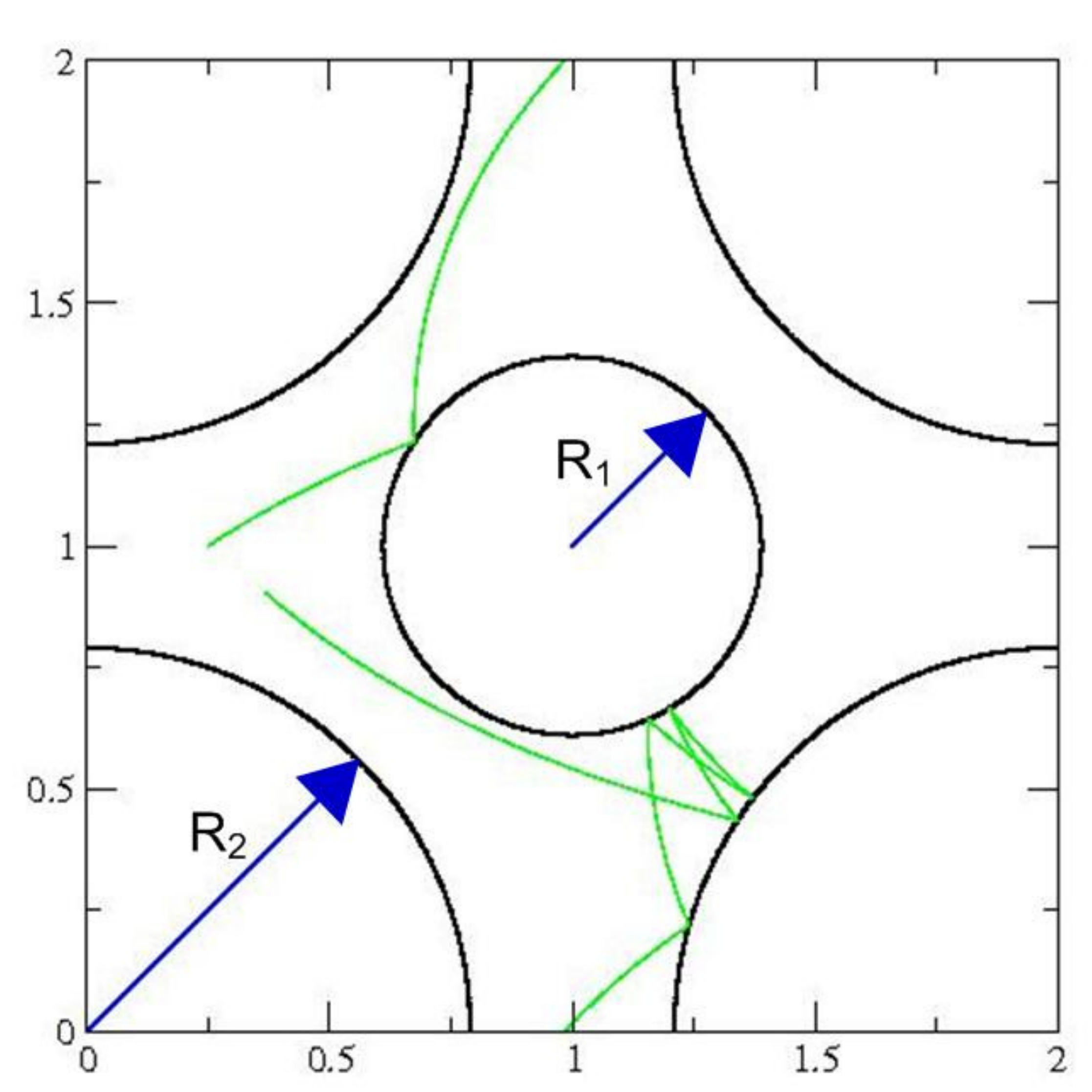}
\caption{{\bf The Sinai billiard}. The radii are $R_1 = 0.39, R_2=0.79$. We
also show an example of trajectory for the external field $\vec
E=(1,0)$.} \label{tres}
\end{center}
\end{figure}

The method of cloning has been shown to work efficiently in the 
verification of the Gallavotti-Cohen theorem on a simple
chaotic system given by the Sinai billiard.
This system consists of a particle moving inside a billiard as in 
figure \ref{tres}, with periodic boundary conditions. 
It is under the action of
a force field $\vec E$, and is subject to a deterministic
thermostat that keeps the velocity modulus constant $|\vec v|=1$.
Between bounces, the equations of motion are:
\begin{eqnarray}
\ddot x_i &=& - E_i +\gamma(t) \dot x_i ,\qquad i=1,2;\nonumber \\
\gamma(t) &=& \sum_i E_i \dot x_i.
\end{eqnarray}
We wish to calculate the fluctuations of the dissipated power $\gamma$ and thus the dynamical partition function
\begin{equation}
Z_t(\alpha)=\langle {\rm e}^{\alpha \int_0^t \gamma(t') dt'}\rangle
\end{equation}
The fluctuation theorem arises from the symmetry 
$$
\mu(\alpha) = \mu(-1-\alpha)
$$
with $\mu(\alpha) = \lim_{t\to\infty} \frac{1}{t} \ln Z_t(\alpha)$. Therefore with reference to the notation 
of the first section we have now
$$
A(x) = \gamma(x)
$$
As in the previous section, the dynamics is deterministic, and hence 
to allow different clones to diversify,  we introduce a small stochastic noise, (cfr paragraph leading to equation \eqref{eqn:hamequabruit}) 
and check the stability of results in the limit of small noise. We evolve the
system for {macroscopic} intervals ${\cal{T}}$, and clone at time $t' = n {\cal T} $
with a factor 
$$
k_{t'}=e^{\alpha \int_{t'}^{t'+{\cal{T}}}\gamma(t'') \,dt''} \;.
$$ 
Before each
deterministic step of time ${\cal{T}}$, clones are given random
kicks of variance $\Delta$ in position  and/or velocity direction.
{The time-interval ${\cal{T}}$ and the noise intensity
$\Delta$ are chosen so that twin clones have a chance to separate
during time ${\cal{T}}$}, and this depends on the chaotic
properties of the system. In the present case, 
$0.1\leq {\cal{T}}\leq 1$  allows for a few collisions, which
guarantees clone diversity for $10^{-3}\leq \Delta \leq 10^{-4}$.

\begin{figure}[ht]
\begin{center}
\includegraphics[width=8.cm]{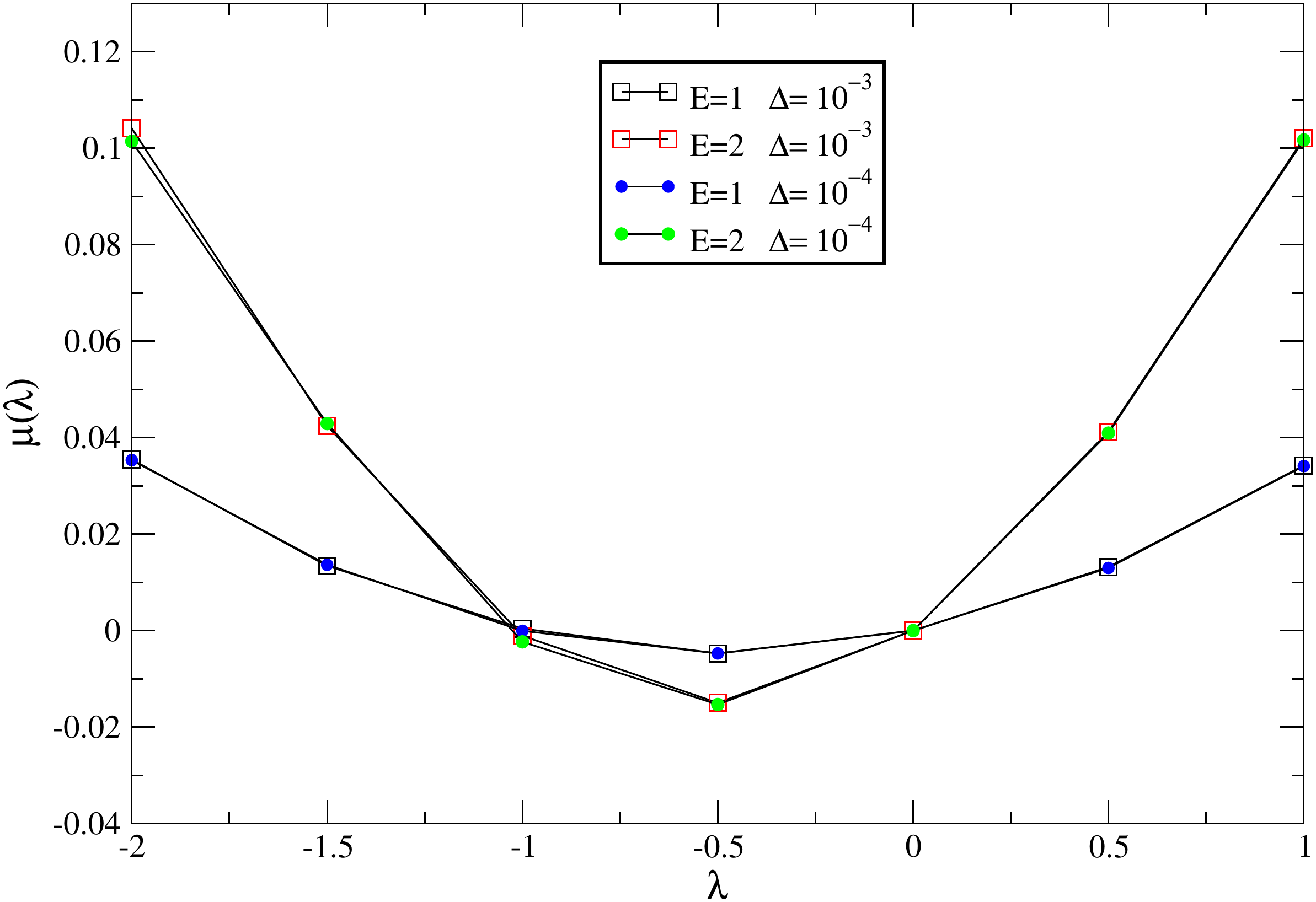}
\end{center}
\caption{{\bf The Gallavotti-Cohen theorem}. Plot of $\mu(\lambda)$ vs.\ $\lambda$ for the driven
Lorentz gas. Data for $\vec E=(E,0)$, $E=1,2$ and noise intensity
$\Delta = 10^{-3},10^{-4}$. The Gallavotti-Cohen theorem implies
the symmetry around $\lambda=-1/2$. The continuous lines
represents a polynomial fit, quadratic for $E=1$ (Gaussian
behavior), 4-th order for $E=2$.} \label{cuatro}
\end{figure}

In Fig.~\ref{cuatro} we show the results of $\mu(\alpha)$
for $-2 \leq \alpha \leq 1$, and for $\vec
E=(E,0)$ with $E=1$ and $E=2$, both corresponding to very large current
deviations  (in the figure $\alpha$ is called $\lambda$).

\section{Planetary systems}
\label{planets}

Planetary systems are the epitome  of  deterministic systems. With their relatively small number of interacting
bodies, they could easily be considered the systems that are further from  statistical treatment.
And yet, statistical analysis of orbits becomes necessary: when we discover  a planetary system we find that many amongst   the  
 observationally allowed configurations  are only stable in the immediate past or future \cite{planet}.
 Since we do not expect that just by chance we came across a system that has just ejected (or will soon eject) a planet,
 we tend to favor amongst  configurations  compatible within error with the data, those that have an unusually high  level
of stability.  

 On a related line, it has been shown \cite{Laskar} that just considering a shift in the Earth's present position of
   the order of one hundred meters,  the fate of Mercury may change dramatically, in some  cases leading its
   orbit to intersect the one of Venus.
Consider for example the study  by Laskar \cite{Laskar}.   
 In a first calculation, he integrated the orbit of Mercury starting from different configurations, obtained
 by displacing the position of the earth by about $150$ {\em meters}.  The orbits obtained this way were qualitatively similar, and yet 
different.
Next, he repeated the calculation but making a few clones of the trajectories, and choosing the one with largest eccentricity. 
After a few such steps, he reached  orbits with great eccentricities, that could cross the orbit of Venus.
We recognize here  a strategy that is very close to the one we are describing here, for the particular cases
$\alpha=0$ and $\alpha=\infty$, respectively. The small displacements are in fact playing the role of our noise.

Indeed, if at each cloning step we had cloned or killed configurations in a fraction proportional  to $\alpha$ times the eccentricity change
during the corresponding time interval 
(cfr.  Section 3:
the eccentricity plays here the role of $\phi$ described there) 
we would have obtained the full probability distribution of, say, the eccentricity at each time. Denote $N(\alpha)$  the total number 
of clones   at time $t$ obtained without normalizing the clone population, or keeping track of the
normalizations if they were done.  $N(\alpha)$ is the Laplace transform of the probability $P(e)$:
\begin{equation}
N(\alpha ) = \int de \; e^{-\alpha e} \; P(e)
\end{equation}
Just as in the example of Sinai's billiard, because the system is chaotic, the displacements (or the noise level),
may be essentially negligible -- for example, compatible with all other external sources of displacements which we have neglected ---
and yet yield all the  variety of  trajectories. 

It would be very interesting to see these methods applied to studying in detail the possible future and past evolution
of planetary  systems, with a large deviation statistical analysis.
Many interesting questions concerning the self-organization of the stability of our solar system could be investigated this way.

\appendix 
\section{Cloning in continuous time: an example pseudo-code }
\label{appendix-pseudo-code}

In this appendix we provide an example pseudo-code for the cloning of
a system described by a configuration \texttt{conf}, evolving with
Markov dynamics in continuous time (see
section~\ref{subsec:continuous-time-cloning}).  The dynamics of each
clone consists in a succession of (i) Poissonian waiting times (sampled
with the function \texttt{random.poisson}) between jumps (ii) change
of configuration, or ``jumps'' (performed by \texttt{evolve()}) and (iii)
cloning, keeping the total number of clone constant. The
way in which the weighted average of a time-extensive observable
\texttt{obs} is computed is also explicited: a value of \texttt{obs}
is attached to each clone and copied/pruned with it.

\begin{footnotesize}
\begin{verbatim}
alpha=0.1          # parameter conjugated to the observable F
N=500              # number of clones
time=0             # initial time
tmax=1000          # maximum simulation time
cloning=0          # logarithm of the global cloning factor 
                   # at the end the ldf is given by cloning/time 
conf.init()        # initialization of the clones:
                   #   conf[1] to conf[N] are set to given configurations
escaperate.init()  # initialization of the alpha-dependent escape rates
obs.init()         # initialize an observable obs that we want to average 
                   # over weighted histories

# initialisation of first jump times
for c from 1 to N do:
  jumptime[c]=random.poisson(escaperate[c])   # Poisson law of rate escaperate[c]

# main loop

while t<tmax do
  (c,t)=next(jumptime)    # returns the first clone c to jump, and its jumptime t
  conf[c].evolve()        # evolves the configurations clone c
                          # note that the observable obs is evolved accordingly
  deltaT=random.poisson(escaperate[c])
                          # determines the time interval until the next jump
  jumptime[c]+=deltaT     # updates the jumptime
  K=conf[c].clfact(deltaT)# yields the cloning factor 
                          #  K=e^(deltaT*(deltaescaperate[c]+alpha*A[c]))
  cloning+=log((N+K-1)/N) # updates the log of the global cloning factor

  k=floor(K+random.real())# integer number k representing the number of clones
                          # replacing the current clone c

  cases
    k=0: # clone c is suppressed, i.e. replaced by another one chosen at random
      do newc=random.integer(N) while newc==c
      conf[c]=conf[newc]
      obs[c]=obs[newc]
      jumptime[c]=jumptime[newc]
    k=1: # nothing is done
    k>1: # k-1 copies of c have to be done; then, among the total N+k-1 
         # resulting clones, k-1 of them are pruned so as to keep N constant
      indices=randomarray(N,k)
         # puts in indices  k-1 *different* random integers between 1 and Nclones+k-1 
         # (both included); only those less or equal than N will be replaced by c
      for newc in indices do:
         if newc<=N do:
            conf[newc]=conf[c]
            obs[newc]=obs[c]
            jumptime[newc]=jumptime[c]

# output of results
ldf=cloning/time
print('large deviation function = ',ldf)
meanobs=sum(obs[c] for c in range(N))/N/time
print('weighted mean of observable = ',meanobs)
\end{verbatim}
\end{footnotesize}

\bibliographystyle{elsarticle-num}

\end{document}